\documentstyle[11pt,paspconf]{article}
\include{epsf}

\begin{document}

\title{Formation of Elliptical and S0 Galaxies by Close Encounters}

\author{Thorsten Naab and Andreas Burkert}
\affil{Max-Planck-Institut f\"ur Astronomie, Heidelberg, Germany}

\begin{abstract}
We study a possible formation mechanism for elliptical/S0 galaxies
using N-body simulations with GRAPE. A close galaxy encounter which
does not lead to a merger can
induce a strong bar in an initially axisymmetric disk. This bar is
unstable to bending oscillations and loses angular momentum to
the dark halo. After 6 Gyrs the remnant is spheroidal with a de
Vaucouleurs-type surface density profile and has ellipticities and rotation
properties comparable to observed elliptical/S0 galaxies: the system rotates
fast $(v/\sigma \approx 1)$ and has disky isophotes.  
\end{abstract}


\section{Introduction}
Most massive elliptical galaxies can be divided into two
groups with different physical properties. Bright ellipticals are slow
anisotropic rotators, have boxy distorted isophotes, are radio-loud,
and are surrounded by gaseous X-ray halos (Bender et al., 1989).
Faint elliptical galaxies are preferentially oblate isotropic
rotators with disky isophotes. In contrast to boxy ellipticals they
are radio-quiet and show no X-ray emission in excess to their discrete
source contribution. Gravitational N-body simulations, starting with
the work by Toomre \& Toomre (1972,1977) using a restricted three-body
approximation, and continued by others (e.g.: Barnes, 1988) using a
hierarchical tree algorithm, have shown that mergers of equal mass
disk-galaxies can produce slowly-rotating remnants with a de
Vaucouleurs-like surface-brightness profile and disky or boxy
isophotes depending on the viewing angle (Heyl et al., 1994).  It is
generally believed that merger remnants are supposed to have
properties resembling observed boxy elliptical galaxies, such as slow
figure rotation, kinematically distinct cores, so the question of how
disky and fast rotating ellipticals have formed is 
still unanswered. Here we study a possible formation mechanism for 
disky elliptical galaxies in an encounter scenario. An approximated massive
galaxy perturbs an equal mass disk galaxy and flies away. We
analyze the triaxial shape, the rotation properties and the isophotal
shape of the remnant.   
\subsection{The model}
The initial conditions for our simulations were derived following
Hernquist (1993) as described briefly below. 
The N-body model consists of an exponential disk surrounded by a dark
halo. The thickness of the disk is determined by the velocity
dispersion which is a function of radius. The halo is initially
spherical and has an isothermal density profile with a core and a
cutoff-radius to reduce 
\newpage
\begin{figure}[!ht]
  \begin{center}
    \leavevmode
    \epsfxsize = 13cm	
    \epsfysize = 17cm	
    \epsffile{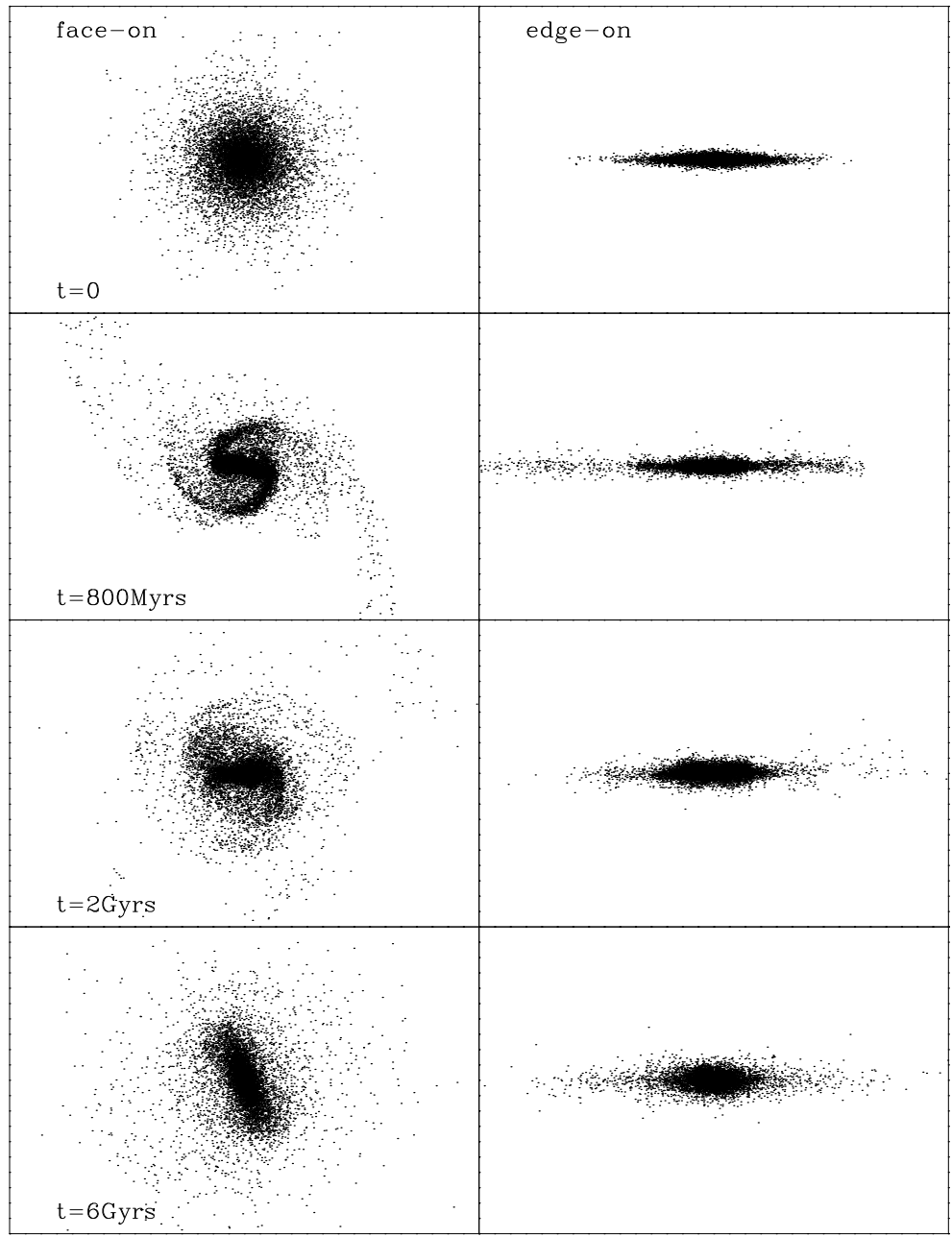}
	   \caption{Snapshots of the simulation at different time
    steps seen face-on and edge-on. Only the distribution of the
    luminous matter is shown.} 
   \end{center}
\end{figure}
the computational costs. Velocities are initialized
by taking moments of the collisionless Boltzmann equation and
approximating the distribution function in phase space by
Gaussians. This produces stable models that are nearly in
equilibrium. The Toomre Q-parameter is normalized to the value of 1.5 at a
radius similar to the Solar neighborhood. The mass and size of the
disk is scaled to physical values of the Milky Way,
i.e. scale length h=3.5 kpc and disk mass $M_d = 5.6 \times
10^{10}_{odot}$. For the perturbing galaxy, we used an equal mass
particle realization of a profile approximating a dark halo. At the
beginning of the simulation the perturber is 
on a prograde hyperbolic orbit with an impact paramter of 63 kpc, and
a relative velocity of $343$ km/s. 
 
The simulation was performed with a direct N-body code using the special
hardware device GRAPE (Sugimoto et al., 1990). We followed the
simulation for 6 Gyrs. The simulation presented here has 50,000 disk
particles, 100,000 halo particles, and 20,000 particles representing the
perturber. 

\subsection{Results} 
Figure 1 shows a sequence of snapshots for our simulation. 
The encounter induces a strong bar in the disk, although the disk is
stable against 
bar formation if simulated in isolation. This bar becomes unstable
to bending oscillations (Raha et al. 1991, Merritt \& Sellwood, 1994)
due to an increase of the velocity dispersion in radial direction. As a
result of this instability the initially disklike system becomes spheroidal.

To estimate the three-dimensional shape of the remnant we computed the
axis ratios of the distribution of disk particles from the eigenvalues of the
moment-of-inertia tensor. For the 60\%  most
tightly bound particles the triaxiality parameter $T \equiv
(a^2-b^2)/(a^2-c^2)$ is 0.88. The effective Hubble types
for projections along the three principal axes are E5.7, E7.3 and E4
respectively.

The surface density follows a de Vaucouleurs-like profile over a large
radial interval and is comparable to the remnants of merger simulations. 
After the strong rotating bar has formed, its dynamical friction with
the live halo component leads to an effective transport of angular momentum to the
halo. The rotation velocity in the inner parts decreases
rapidly. Figure 2 shows the rotation properties of the system after
6 Gyrs. We  have plotted $v_m/\sigma$ ($v_m$: maximum rotation
velocity, $sigma$: central velocity dispersion)  against the ellipticity. The
points show the values for the same remnant in different
projections. One can see that some projections follow the line for 
oblate rotators, but we also have anisotropic remnants and those with 
very high rotation velocities at low ellipticities. This effect can 
be influenced by the determination of the ellipticity; some observers
see the  same effect in their data (see Nieto et al., 1988). 

We also investigated the deviations of the isophotes from pure
ellipses applying the method described by Bender el al. (1988) after
binning the particle distribution and convolving it with a Gaussian with
a FWHM comparable to the seeing conditions of the observations. 

We find that the remnant has disky isophotes (positive $a4/a$)  
for almost all projections. The value of $a4$ is higher for remnants with
higher ellipticity (Figure 2). With increasing radius the $a4$
coefficient shows basically two global features: either a disky inner
part changing to 
boxy in the outer parts, or a continuously rising positive $a4$. Those
features could be explained as suggested before (Nieto et al., 1991) by a faint
disk surrounded by a spheroidal component or, in the case of rising
$a4$, tidal extensions of a round inner part. In our simulation the shape of
the profile just depends on the projection angle. 

\begin{figure}[!ht]
  \begin{center}
    \leavevmode
    \epsfxsize = 14cm	
    \epsfysize = 7cm	
    \epsffile{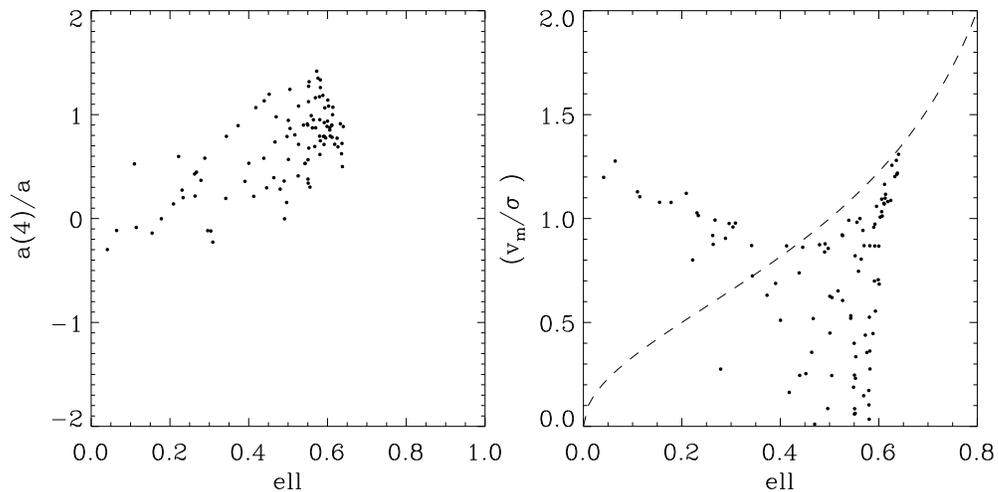}
	   \caption{Left: The isophote-shape parameter a(4)/a against
    ellipticities for 100 randomly chosen projections (positive
    a(4)/a: disky, negative a(4)/a: boxy). Right: $v_m /
    \sigma$ against ellipticity. } 
   \end{center}
\end{figure}

Future simulations will show how sensitive our results are to
the resolution and different initial conditions.


\begin{references}
\reference Barnes , J., E. 1988, \apj, 331, 699
\reference Bender, R. ,D\"obereiner, S. \& M\"ollenhoff, C. 1988,
\aaps, 74, 385  
\reference Bender, R., Surma, P. ,D\"obereiner, S., M\"ollenhoff,
C. \& Madejsky, R. 1989, \aap, 217, 35  
\reference Davies, L. ,D. , Efstathiou, G., Fall, S., M., Illingworth,
G. \& Schechter, P. 1983, \apj, 266, 41
\reference Hernquist, L. 1993, \apjsupp, 86, 389
\reference Heyl, J., S., Hernquist, L.\& Spergel, D. N. 1994, \apj,
463, 69 \
\reference Nieto, J.-L., Capaccioli, M., Held, E. V. 1988, \aap,
195, L1 
\reference Nieto, J.-L., Poulain, P., Davoust, E. \& Rosenblatt,
P. 1991, \aaps, 88, 559 
\reference Merritt, D., Sellwod, J. 1994, \apj, 425, 551
\reference Raha, N., Sellwood, J., A., James, R., A. \& Kahn,
F. D. 1991, Nature, 352, 411
\reference Sugimoto, D., Chikada, Y., Makino, J., Ito, T., Ebisuzaki,
T.\&  Umemura, M. 1990, Nature, 345, 33
\reference Toomre, A \& Toomre, J. 1972, \apj, 178, 623
\reference Toomre, A 1977, in The Evolution of Galaxies and Stellar
Populations, ed. B. Tinsley\& R.Larson (New Haven: Yale
Univ. Press),p. 401
\end{references}
\end{document}